\documentclass[
amsmath,
preprint,
pra
]{revtex4}

\usepackage{braket}
\usepackage{mathrsfs}
\usepackage{graphicx}

\graphicspath{{pics/}}

\def\rhohat{\hat{\rho}}
\def\gamhat{\hat{\gamma}}

\def\Xhatplus{\hat{X}^+}
\def\Xhatminus{\hat{X}^-}

\begin{document}

\title{Non-Gaussian, Mixed Continuous-Variable Entangled States}
\author{A. P. Lund}
\email{lund@physics.uq.edu.au}
\author{T. C. Ralph}
\affiliation{Centre for Quantum Computer Technology, Department of Physics,\\
University of Queensland, St Lucia, QLD 4072, Australia}

\author{P. van Loock}
\affiliation{National Institute of Informatics,\\
Chiyoda, Tokyo 101-8430, Japan}

\begin{abstract} 

  We study a class of mixed non-Gaussian entangled states that, whilst
  closely related to Gaussian entangled states, none-the-less exhibit
  distinct properties previously only associated with more exotic,
  pure non-Gaussian states.

\end{abstract}
\maketitle

\section{Introduction}

States with non-Gaussian quadrature probability distributions are of
high interest in quantum optics.  Pure versions of such states are
difficult to generate experimentally as they require either very high
non-linearity \cite{turchette:non-gaussian} or difficult conditional
preparation \cite{lvovsky:single-photon}. If we include multi-mode and
mixed non-Gaussian states we open up a huge arena that is in general
not fully understood. In contrast quantum states of light which have
Gaussian quadrature distributions are well understood \cite{walls:qo}.
This now includes their entanglement properties
\cite{duan:inseparability}. Numerous quantum information applications
for Gaussian states have been proposed \cite{braunstein:CVQI} and demonstrated
\cite{furusawa:1998}, \cite{grosshans:gaussianQKD}, \cite{lance:2004}, \cite{takei:2005}.

Here we propose an entangled state which is non-Gaussian but could be
considered one step away from a Gaussian state.  The state is a random
mixture of two distinct Gaussian states.  This is quite clearly
non-Gaussian as adding two Gaussian distributions with different means
or different variances will result in either a ``two peaked''
distribution or a distribution where the higher order moments are not
what is expected from a Gaussian distribuiton.  So-called "proper
mixtures" of this kind, i.e. where the mixture is created by
introducing a classical random variable, have been studied extensively
in the discrete variable domain of photon counting experiments
\cite{munro:2001}, \cite{wei:2005}. We show that our proposed continuous
variable mixed states exhibit interesting properties that have
previously only been observed in the context of models of more exotic,
pure non-Gaussian states. Here the effects are found in a more
accessible scenario.



We begin in section~\ref{measures} by introducing the measures of
entanglement that we will use for our analysis. Section~\ref{gaussian}
will discuss the entanglement properties of Gaussian states and
section~\ref{mixtures} will extend this to our mixed states.  In
section~\ref{examples} 
we discuss two examples: continuous variable quantum teleportation
\cite{furusawa:1998} and; the entanglement cloner attack in coherent
state quantum key distribution \cite{grosshans:gaussianQKD}. Finally
in section~\ref{conclusion} we will conclude.

\section{Entanglement Measures}
\label{measures}

In this paper we will require some way to compare entangled states.
The measures used here are the so called
`negativity'~\cite{vidal:computable_entanglement} and the
`inseparability criterion'~\cite{duan:inseparability}.  These measures contain
most of the properties that one would expect from a measure of
entanglement.  We choose only these two here as they are comparitively
easy to compute in the continouous variable regime.

\subsection{Negativity}

We will first introduce the
negativity~\cite{vidal:computable_entanglement}.  This entanglement
measure is the sum of the negative eigenvalues of the partial
transpose of the density operator~\cite{peres:separability}.
In~\cite{vidal:computable_entanglement} the properties of this measure
are studied and it is shown that this value can be computed by
evaluating
\begin{equation}
\label{negativity}
\mathscr{N}(\hat{\rho}) = \frac{\lVert \hat{\rho}^{T_A} \rVert_1 - 1}{2},
\end{equation}
where $\lVert \cdot \rVert_1$ is the usual trace norm (i.e. $tr
\sqrt{\hat{A}^\dagger \hat{A}}$) and $\hat{\rho}^{T_A}$ is the partial
transpose of the density operator~\cite{peres:separability}.

\subsection{Inseparability criterion}

The `inseparability criterion'~\cite{duan:inseparability} using the
classic EPR type observables~\cite{einstein:EPR} provides a value
which can be used as a sufficient condition for inseparability.  This
is useful as it gives a way of knowing if a state is entangled by
calculating values from averages of obserable quantities.  
The
inseparability criterion also has an operational meaning in CV
teleportation.  There is a one-to-one correspondance between it and
the fidelity of teleportation if the entanglement resource has a
symmetric Gaussian Wigner function.  This is an important point for
this work, as later we will use this fact to argue that the
inseparability criterion is a good quantity to use to compare
continuous variable entangled states.

If we choose our energy scale so that shot noise has unit variance then
the inseparability criterion can be written
\begin{equation} 
\label{inseparability}
\frac{1}{4}\langle \left[\Delta (\Xhatplus_1 + \Xhatplus_2)\right]^2 \rangle +
\frac{1}{4}\langle \left[\Delta (\Xhatminus_1 - \Xhatminus_2)\right]^2 \rangle  < 1
\end{equation}
where the subscripts $1$ and $2$ represent the two modes in which the
entanglement exists and $\hat{X}^+_j = \hat{a}_j +
\hat{a}^{\dagger}_j$, $\hat{X}^+_j = i(\hat{a}_j -
\hat{a}^{\dagger}_j)$ are the in-phase and out-of-phase quadrature
amplitudes respectively. Also $\hat{a}_j$ and $\hat{a}_j^{\dagger}$
are the anhillation and creation operators respectively for the modes.
Note that this expression is only valid when considering entanglement
that has symmetric correlations, both between the modes and between
their quadartures (technically, the entanglement is said to be in the
standard form \cite{duan:inseparability}). The expression on the left
hand side of equation~\ref{inseparability} we shall call the
`inseparability criterion number' or just inseparablity criterion when
the meaning obviously implies a numerical value as opposed to an
inequality.

\section{Two-mode Gaussian states}
\label{gaussian}


The Gaussian states we will use as ingredients for our non-Gaussian
mixed states are the two mode squeezed vacua. We can write
\begin{equation}
\label{sqz_vaccum}
\hat{S}(r) \ket{0,0} = e^{r(\hat{a}_1 \hat{a}_2 - \hat{a}_1^\dagger 
\hat{a}_2^\dagger)} \ket{0,0}
\end{equation}
where $\hat{S}(r)$ is the `squeeze' operator with $r$ as a parameter
(here assumed real) which determines the amount by which the state is
squeezed (squeeze parameter). This state is Gaussian and hence is
completely characterized by a vector of means and a covariance matrix.
A two mode state requires four phase space coordinates and we will
choose to write them in a vector of the form $(x_1,p_1,x_2,p_2)$.
Squeezed vaccua are centered around the origin of phase space, so the
vector of means is the zero vector.  The covariance matrix for the two
mode squeezed vacuum as defined in Eq.~\ref{sqz_vaccum} is
\begin{equation}
\begin{pmatrix}
\cosh(2r) & 0 & -\sinh(2r) & 0 \\
0 & \cosh(2r) & 0 & \sinh(2r) \\
-\sinh(2r) & 0 & \cosh(2r) & 0 \\
0 & \sinh(2r) & 0 & \cosh(2r) 
\end{pmatrix}.
\end{equation}
Given these variances we are now in a position to evaluate the
inseparability criterion for this state.  It is
\begin{equation}
\label{gaussian_inseparability}
\cosh(2r) - \sinh(2r) = e^{-2r}.
\end{equation}
Note that this value is less than one and hence is entangled for all
non-zero, positive $r$.

The two mode squeezed vaccum when written in the Fock basis is
\begin{equation}
\hat{S}(r)\ket{0,0} = \sum_{n=0}^\infty (\tanh r)^n \ket{n,n}.
\end{equation}
Using this basis, one can show that the negativity of this state is
\begin{equation}
\label{gaussian_negativity}
\mathscr{N}(\hat{\gamma}(r)) = \frac{\tanh r}{1-\tanh r} = \frac{e^{2r}-1}{2}.
\end{equation}

Using this state as the entanglement resource to teleport an unknown
coherent state using continuous variable
teleportation~\cite{braunstein:contvarteleportation} (with unity gain)
results in an output state which when compared with the input state
has a fidelity of
\begin{equation}
\mathscr{F}_{\hat{S}(r) \ket{0,0}} = \frac{1}{1+e^{-2r}}.
\end{equation}
Note here that the fidelity is in one-to-one correspondance with the
inseparability criterion.  This is only true for symmetric Gaussian
states.

\section{Mixtures of Gaussian states}
\label{mixtures}

Consider the mixed state which is a random mixture of vaccum state and
two mode squeezed state, i.e.
\begin{equation} 
\label{vaccum_mixture}
\rhohat_{rm1} = p \gamhat(r) + (1-p) \gamhat(0) 
\end{equation}
where $\gamhat(r)$ represents the density operator associated with a
Gaussian two mode squeezed vaccum with squeeze parameter $r$ and hence
$p$ is the probability that the state is actually the \emph{squeezed}
vaccum.  Note here that $\gamhat(0)$ is the un-squeezed vaccum and
hence is the usual vaccum state.

To calculate the inseparability criterion for this state involves
taking the expectation value of given operators (see
Eq.~(\ref{inseparability})).  Expectation values are calculated in
quantum mechanics by calculating $\langle \hat{A} \rangle = Tr \left\{
  \hat{\rho} \hat{A} \right\}$.  This equation is linear in both
$\hat{A}$ and $\hat{\rho}$.  So the inseparability criterion reduces
to evaluating the weighted sum of the inseparability criterions of the
two states which make up the mixture.  That is
\begin{equation}
\label{vaccum_mixture_inseparability}
p e^{-2 r} + (1 - p).
\end{equation}
The negativity of the mixture is given by
\begin{equation}
\label{non-gaussian_negativity}
p(\frac{e^{2r}-1}{2}).
\end{equation}

\section{Comparison}

One way to compare the Gaussian state with the non-Gaussian state
would be to compare the two states with the same $r$ (squeezing). This
type of comparison was made by Mista et al in the context of similar,
thermal/squeezed state mixtures ~\cite{mista:cv_werner}. It is clear
that with such a criteria the non-Gaussian state will always be
inferior to the Gaussian state.  However, we believe that this is not
a fair comparison of the two states.

Our method for comparing these two states is by way of the
inseparability criterion.  We take this stance as the inseparability
criterion is generally accepted as a measure of the "strength" of
entanglement for Gaussian states. Operationally, as we have seen, it
is in one-to-one correspondence with the fidelity of CV teleportation
of coherent states and is easily measured via the second order
moments.  We are interested in ways in which our non-Gaussian state
may behave differently from Gaussian states, thus it makes sense to
compare states with the same "strength" of entanglement, i.e. the same
inseparability criterion.

If we take a mixture of two Gaussian states as per
Eq.~(\ref{vaccum_mixture}) then a pure two mode squeezed state would
have the same inseparability criterion when
\begin{equation}
\label{fair_comp}
e^{-2 s} = p e^{-2 r} + (1 - p) = I
\end{equation}
where $s$ is the squeeze parameter for the pure state.  Here we are
defining $I$ to be the common inseparability criterion.  As $0 < e^{-2
  r} \le 1$ and $0 \le p \le 1$ we find that $1-p < I \le 1$.  It is
important to note here that as the squeezing increases, $I$ decreases.
So the lowest possible $I$ corresponds to the maximum squeezing.

\subsection{Negativity}

\begin{figure}
\includegraphics[width=8cm]{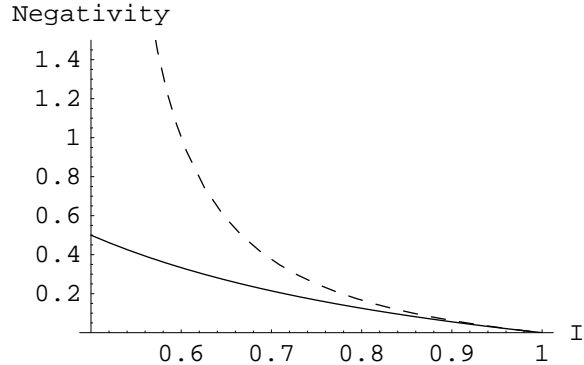}
\caption{A plot of the negativity of the mixed state as per
Eq.~(\ref{negativity_mixed}) (dashed) and the equivalent pure state
with the same inseparability criterion as per
Eq.~(\ref{negativity_pure}) (solid).  This plot is in the same style
as Fig.~\ref{fair_comp_fidelity_plot}.}
\label{negativity_comp}
\end{figure}

The negativity of the non-Gaussian state of Eq.~(\ref{vaccum_mixture})
can be expressed in terms of the $I$ using Eq.~(\ref{fair_comp}) as
\begin{equation}
\label{negativity_mixed}
\mathscr{N}(\hat{\rho}_{rm1}) = \frac{p^2}{2(I+p-1)} - \frac{p}{2}.
\end{equation} 
Similarly the negativity of the Gaussian state can be written
\begin{equation}
\label{negativity_pure}
\mathscr{N}(\hat{\gamma}(s)) = \frac{I^{-1}-1}{2}.
\end{equation}
The Eqs.~(\ref{negativity_mixed}) and~(\ref{negativity_pure}) are
plotted for comparison in Fig.~\ref{negativity_comp}.
It would appear that the negativity of the non-Gaussian state is
always higher than that of a Gaussian state with the same
inseparability criterion.  To prove this we
consider the difference between the mixed state negativity
(Eq.~(\ref{negativity_mixed})) and the pure state negativity
(Eq.~(\ref{negativity_pure})):
\begin{equation}
\frac{p^2}{2(I+p-1)}-\frac{p}{2}-\frac{I^{-1}-1}{2} = \\
\frac{(I-1)^2 (p-1)}{2I(I-1+p)},
\end{equation}
which is always a positive quantity as $0 \le p \le 1$ and $1-p < I \le 1$.

We conclude that the entanglement strength as assessed by the trace
negativity is always greater for our non-Gaussian states even though
they exhibit the same entanglement strength as evaluated using the
inseparability criterion. Next we examine an operational consequence
of this fact.

\section{Two examples}
\label{examples}

With the knowledge from the previous section, we are able so show how
these results apply to two practial situations where one may use this
non-Gaussian state.  The examples we will give involve using the
non-Gaussian state as the entanglement for continuous variable quantum
teleportation and the entanglement cloner attack in coherent state
quantum key distribution.

\subsection{Teleportation}

It has previously been noted that pure non-Gaussian states, if used as
a resource in CV teleportation, can teleport coherent states with a
higher fidelity than Gaussian resource states with the same
inseparability criterion \cite{cochrane:2002}. We now demonstrate that this
unusual property is also possessed by our mixed non-Gaussian states.

\begin{figure}
\includegraphics[width=8cm]{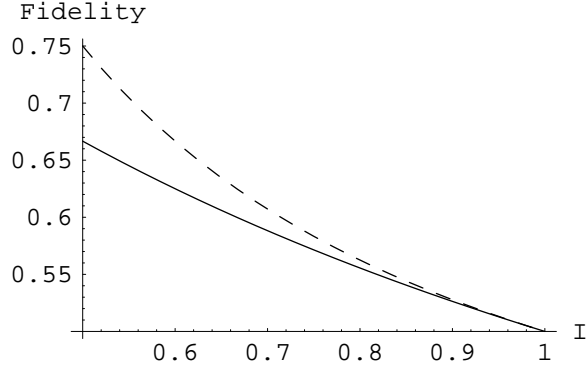} 
\caption{Plot of the pure state fidelity as per
Eq.~(\ref{pure_comp_fidelity}) (solid line) and mixed state fidelity as per
Eq.~(\ref{vaccum_mixture_fidelity}) (dashed line) for $p=0.5$ as a function
of $r$.  Note that as the squeezing increases $I$ decreases.}
\label{fair_comp_fidelity_plot}
\end{figure}

The fidelity of continuous variable teleportation performed using our
non-Gaussian state as the entanglement resource is simply given by the
weighted sum of the fidelities achieved by its two constituent
Gaussian states,
\begin{equation}
\label{vaccum_mixture_fidelity}
\mathscr{F}_{\rhohat_{rm1}} = \frac{p}{1+e^{-2r}} + \frac{1-p}{2}.
\end{equation}
The fidelity of CV teleportation when using Gaussian state with squeeze
parameter given by Eq.~(\ref{fair_comp}) is
\begin{equation}
\label{pure_comp_fidelity}
\frac{1}{1+e^{-2 s}} = \frac{1}{1+I}
\end{equation}
and rewriting Eq.~(\ref{vaccum_mixture_fidelity}) using
Eq.~(\ref{fair_comp}) the non-Gaussian state fidelity is
\begin{equation}
\label{fair_comp_fidelity}
\frac{p}{1+e^{-2r}} + \frac{1-p}{2} = \frac{p^2}{I+2p-1} + \frac{1-p}{2}.
\end{equation}
Fig.~\ref{fair_comp_fidelity_plot} shows this fidelity as a function
of $I$ compared to the fidelity given by
Eq.~(\ref{vaccum_mixture_fidelity}) when $p = 0.5$.  The fidelity for
the mixed state is higher for all values of $r$.
At this point we now claim that this feature, that the non-Gaussian
resource state gives higher fidelity, holds for all $p$ and all $r$ as
long as the inseparability comparison is satisfied.  To show this,
first write down the difference of the mixed state fidelity
(Eq.~(\ref{fair_comp_fidelity})) from the pure state fidelity
(Eq.~(\ref{pure_comp_fidelity})) with the same inseparability
criterion
\begin{equation}
\frac{p^2}{I+2p-1}+\frac{1-p}{2}-\frac{1}{1+I} = \\ 
\frac{(I-1)^2 (1-p)}{2 (I+1)(I-1+2p)}.
\end{equation} 
Note that as $0 \le p \le 1$ and $1-p < I \le 1$ the term on the right
hand side is non-negative.  Hence it follows that
\begin{equation}
\frac{p^2}{I+2p-1}+\frac{1-p}{2} \ge \frac{1}{1+I}.
\end{equation}  
The difference between the two fidelities is maximised when $I = 1-p$
(i.e.  $e^{-2 r} = 0$, infinite squeezing in the mixed state) and $p =
2 - \sqrt{2}$.  The maximum difference in Fidelity is $\frac{1}{2}
\left(3 - 2 \sqrt{2} \right) \approx 8.6\%$.



\subsection{Continuous Variable QKD}

The optimum individual eavesdropper attack on continuous variable,
coherent state, quantum key distribution has been shown to be via an
entangling cloner 
\cite{grosshans:gaussianQKD}. 
One might think that because of the
increased entanglement of the non-Gaussian source, that it would be a
superior resource for implementing such a cloner. On the other hand
Grosshans
\cite{grosshans:gaussianQKDoptimal} 
has shown, in a general way, that Gaussian attacks are optimal. Our
mixed non-Gaussian source offers an accessible example of Grosshans'
general result.



The quantity which determines the rate at which the two main parties
can form a key via the reverse reconsiliation protocol is given by the
difference between their mutual information and the mutual information
between an evesdropper and the recieving
station~\cite{grosshans:gaussianQKDoptimal}. That is 
\begin{equation}
\Delta I = I(B;A) - I(B;E)
\end{equation}
where $I(B;A) = H(Q_B) - H(Q_B|Q_A)$ and $H(\cdot)$ is the entropy of the
measured data $Q$ and $H(\cdot|\cdot)$ is the mutual entropy between the two
parties.  The mutual information for the protocol described
in~\cite{grosshans:gaussianQKD} can be written as
\begin{equation}
\Delta I (A, \eta, N) = \frac{1}{2} \log_2\left(\frac{\left(\frac{\eta}{A} + (1-\eta) N\right)^{-1}}{\eta + (1-\eta) N} \right)
\end{equation}
where $\eta$ is the effiency of the channel and Alice modulates the
input with variance $A$ and $(1-\eta) N$ is the variance of the noise
added by the the channel due to thermal effects.

If now one assumes that an eavesdropper has full control over the
channel, one must make the conservative assumtion that she can control
and manipulated all aspects of the channel.  Here we have a channel
with a given loss rate $\eta$, which has presumably been characterised
by Alice and Bob.  Eve now has access to the input state to the loss
modes of the channel.  Presumably, Alice and Bob would notice if the
loss modes contained a state significantly different from a thermal
state when they characterise the channel.  So the eavesdropper must
try to maintain a state which has similar properties to the thermal
state that Alice and Bob are expecting.

Here we are going to assume that Alice and Bob only characterise their
channel by computing the first and second order moments of the data
they collect from the channel.  They do this because they are
confident that the channel is Gaussian.  The eavesdropper may try to
exploit this by choosing to use her entangling cloner to attach a
mixed state of the form in equation~\ref{vaccum_mixture} such that
\begin{equation}
N = p N_p + (1-p)
\end{equation}
where $N$ is the variance of the state entering the loss mode of the
channel, $N_p$ is the single mode variance from the non-vaccum part of
the state in equation~\ref{vaccum_mixture} and $p$ is as per the same
equation.  

\begin{figure}
\includegraphics[width=8cm]{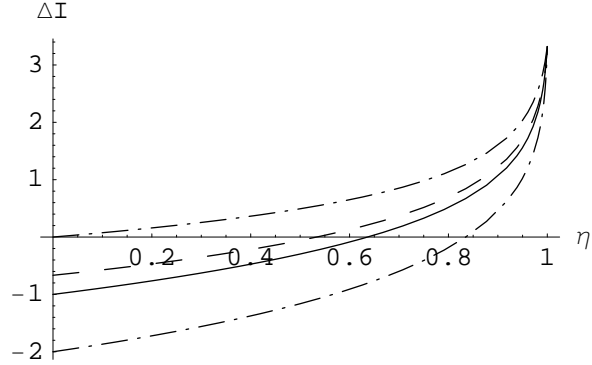}
\caption{A plot of the different of information rates between Alice
  and Bob and Eve and Bob as a function of channel loss.  The solid
  line is where an eavesdropper uses a gaussian state in her
  entangling cloner.  The dashed line is where she uses a mixed state
  in equation~\ref{vaccum_mixture}.  The dot dashed lines are the
  rates that would be achieved if only the components of the mixed
  state were used.  The vaccum state is the top dot-dashed line and
  the squeezed state is the lower dot-dashed line.  Since these two
  states are mixed to produce a state with the same first and second
  order moments as that used in the solid line (see text), the
  resultant differerence in information rate given by the dashed line
  only disadvantages Eve in her attack.}
\label{information-rates}
\end{figure}

Given this restriction on the eavesdroppers state which will decieve
Alice and Bob and using the knowledge that Alice and Bob don't have
about the true nature of the channel, it is possible to calculate
the true difference in mutual information between Alice and Bob and
Eve and Bob as
\begin{equation}
\Delta I_{mix} (A,\eta,N,N_p) = \frac{N-1}{N_p-1} \Delta I(A,\eta,N_p) + (1-\frac{N-1}{N_p-1}) \Delta I (A,\eta,1).
\end{equation}
The eavesdropper's objectives is to minimise the mutual information
between Alice and Bob and maximise that between herself and Bob, in
effect reducing $\Delta I$.  She will achieve this when
\begin{equation}
\Delta I (A,\eta,N) - \Delta I_{mix} (A,\eta,N,N_p) > 0.
\end{equation}
However, an exhaustive numerical search reveals no parameters
$A$,$\eta$,$N$ and $N_P$ satisfying $A,N,N_p>1$,$0<\eta<1$ and $N_p >
N$ which solve this inequality.  A particular example of this is shown
in Figure~\ref{information-rates}.

Hence even though the mixed non-Gaussian entanglement has higher
entanglement and can produce higher fidelity in continious variable
teleportation than the corresponding pure Gaussian entanglement, its
use in an entangleing cloner attack on a continouous variables QKD
protocol does not actually achieve better eavesdropping.

\section{Discrete versus Continuous Loss}
\label{loss}

An objection that could be raised regarding our comparison technique
is that we are comparing states that in general have different photon
numbers and thus have different entanglement per photon. It could then
be argued that a fairer comparison would be to keep both $I$ and the
average photon number the same. We will now show that this scenario
leads to identical conclusions to those before.

Consider the following situation. Greg and Nancy both start with
identical two-mode squeezed states. Both reduce the average photon
numbers in their states by the same ratio but via different
processes. Greg subjects his beams to equal levels of continuous loss,
such as is induced by passage through beamsplitters. Nancy instead
subjects her beams to discrete loss by randomly either completely
blocking or completely transmitting both her beams. If the
transmission efficiency of Greg's state through the continuous loss,
$\eta$ is equal to the probability that Nancy's state is allowed to be
transmitted, $p$, then the average photon numbers of the the two
states after the loss processes are equal. Moreover we will now show
that $I$ is also equal for the two states. However, Greg's state has
remained Gaussian whilst Nancy's state has become non-Gaussian.

Greg's state after the continuous loss is a two mode
Gaussian state with mean zero and covariance matrix given by
\begin{widetext}
\begin{equation}
\begin{pmatrix}
\eta \cosh(2r) + 1 - \eta & 0 & -\eta \sinh(2r) & 0 \\
0 & \eta \cosh(2r) + 1 - \eta & 0 & \eta \sinh(2r) \\
-\eta \sinh(2r) & 0 & \eta \cosh(2r) + 1 - \eta & 0 \\
0 & \eta \sinh(2r) & 0 & \eta \cosh(2r) + 1 - \eta  
\end{pmatrix}.
\end{equation}
\end{widetext}
Note that this covariance matrix is idential to that of the state
given in Eq.~(\ref{vaccum_mixture}) when $\eta = p$. In other words it
has the same covariance matrix as Nancy's state.  Because of this the
inseparability criterion is the same for Greg and Nancy's states.  The
negativity for Greg's state is given by
\begin{equation}
\frac{(1+\eta(e^{-2r}-1))^{-1}-1}{2}.
\end{equation}
Using Eq.~(\ref{fair_comp}) to rewrite this in terms of the
inseparabiltiy criterion and $\eta = p$ gives the negativity to be
\begin{equation}
\frac{I^{-1} - 1}{2},
\end{equation}
which is the same as the equation for the pure Gaussian state
negativity, Eq.~(\ref{negativity_pure}).  So the comparison between
Greg and Nancy's state negativities is identical to that already
plotted in Fig.~\ref{negativity_comp}, i.e. the non-Gaussian state
resulting from discrete loss is always more entangled.

\section{Conclusion}
\label{conclusion}

In this paper we have discussed the non-Gaussian mixed state formed
from the random mixture of a two mode squeezed vaccum and a standard
vaccum state. We have shown that this state exhibits higher
entanglement than an equivalent Gaussian state, where by equivalent we
mean that both exhibit the same level of second order correlation as
measured by the inseparability criterion. From a more operational
point of view we have shown that our non-Gaussian state, if used as a
resource state for continuous variable teleportation, can achieve
fidelities superior to that of an equivalent Gaussian resource
state. Such properties have previously been observed for pure
non-Gaussian entangled states. One might have speculated that these
were "deep" quantum effects, arising from Wigner function
negativity. Our mixed non-Gaussian states do not exhibit Wigner
function negativity thus negating this speculation.

We have used our non-Gaussian state as an example to illustrate the
optimality of Gaussian attacks in continuous variable quantum key
distribution. We showed explicitly that even though the non-Gaussian
state is more entangled, and even if Alice and Bob blindly assume that
the attack is Gaussian, it does not lead to an advantage for Eve. We
have also used our results to study the very distinct characteristics
of discrete and continuous loss on entanglement.

Although not trivial, the production of our suggested states are
within current technological capabilities.

\begin{acknowledgments}
The authors would like to thank Henry Haselgrove for useful suggestions.
\end{acknowledgments}

\bibliography{thebib}

\begin{thebibliography}{18}
\expandafter\ifx\csname natexlab\endcsname\relax\def\natexlab#1{#1}\fi
\expandafter\ifx\csname bibnamefont\endcsname\relax
  \def\bibnamefont#1{#1}\fi
\expandafter\ifx\csname bibfnamefont\endcsname\relax
  \def\bibfnamefont#1{#1}\fi
\expandafter\ifx\csname citenamefont\endcsname\relax
  \def\citenamefont#1{#1}\fi
\expandafter\ifx\csname url\endcsname\relax
  \def\url#1{\texttt{#1}}\fi
\expandafter\ifx\csname urlprefix\endcsname\relax\def\urlprefix{URL }\fi
\providecommand{\bibinfo}[2]{#2}
\providecommand{\eprint}[2][]{\url{#2}}

\bibitem[{\citenamefont{Turchette et~al.}(1995)\citenamefont{Turchette, Hood,
  Lange, Mabuchi, and Kimble}}]{turchette:non-gaussian}
\bibinfo{author}{\bibfnamefont{Q.~A.} \bibnamefont{Turchette}},
  \bibinfo{author}{\bibfnamefont{C.~J.} \bibnamefont{Hood}},
  \bibinfo{author}{\bibfnamefont{W.}~\bibnamefont{Lange}},
  \bibinfo{author}{\bibfnamefont{H.}~\bibnamefont{Mabuchi}}, \bibnamefont{and}
  \bibinfo{author}{\bibfnamefont{H.~J.} \bibnamefont{Kimble}},
  \bibinfo{journal}{Physical Review Letters} \textbf{\bibinfo{volume}{75}},
  \bibinfo{pages}{4710} (\bibinfo{year}{1995}),
  \urlprefix\url{http://link.aps.org/abstract/PRL/v75/p4710}.

\bibitem[{\citenamefont{Lvovsky et~al.}(2001)\citenamefont{Lvovsky, Hansen,
  Aichele, Benson, Mlynek, and Schiller}}]{lvovsky:single-photon}
\bibinfo{author}{\bibfnamefont{A.~I.} \bibnamefont{Lvovsky}},
  \bibinfo{author}{\bibfnamefont{H.}~\bibnamefont{Hansen}},
  \bibinfo{author}{\bibfnamefont{T.}~\bibnamefont{Aichele}},
  \bibinfo{author}{\bibfnamefont{O.}~\bibnamefont{Benson}},
  \bibinfo{author}{\bibfnamefont{J.}~\bibnamefont{Mlynek}}, \bibnamefont{and}
  \bibinfo{author}{\bibfnamefont{S.}~\bibnamefont{Schiller}},
  \bibinfo{journal}{Physical Review Letters} \textbf{\bibinfo{volume}{87}},
  \bibinfo{pages}{050402} (\bibinfo{year}{2001}).

\bibitem[{\citenamefont{Walls and Milburn}(1994)}]{walls:qo}
\bibinfo{author}{\bibfnamefont{D.~F.} \bibnamefont{Walls}} \bibnamefont{and}
  \bibinfo{author}{\bibfnamefont{G.~J.} \bibnamefont{Milburn}},
  \emph{\bibinfo{title}{Quantum Optics}} (\bibinfo{publisher}{Cambridge
  University Press}, \bibinfo{year}{1994}).

\bibitem[{\citenamefont{Duan et~al.}(2000)\citenamefont{Duan, Giedke, Cirac,
  and Zoller}}]{duan:inseparability}
\bibinfo{author}{\bibfnamefont{L.-M.} \bibnamefont{Duan}},
  \bibinfo{author}{\bibfnamefont{G.}~\bibnamefont{Giedke}},
  \bibinfo{author}{\bibfnamefont{J.~I.} \bibnamefont{Cirac}}, \bibnamefont{and}
  \bibinfo{author}{\bibfnamefont{P.}~\bibnamefont{Zoller}},
  \bibinfo{journal}{Physical Review Letters} \textbf{\bibinfo{volume}{84}},
  \bibinfo{pages}{2722} (\bibinfo{year}{2000}),
  \urlprefix\url{http://link.aps.org/abstract/PRL/v84/p2722}.

\bibitem[{\citenamefont{Braunstein and Pati}(2003)}]{braunstein:CVQI}
\bibinfo{editor}{\bibfnamefont{S.}~\bibnamefont{Braunstein}} \bibnamefont{and}
  \bibinfo{editor}{\bibfnamefont{A.}~\bibnamefont{Pati}}, eds.,
  \emph{\bibinfo{title}{Continuous Variable Quantum Information}}
  (\bibinfo{publisher}{Kluwer Academic Publishers}, \bibinfo{year}{2003}).

\bibitem[{\citenamefont{Furusawa et~al.}(1998)\citenamefont{Furusawa,
  S\"{o}rensen, Braunstein, Fuchs, Kimble, and E.S.Polzik}}]{furusawa:1998}
\bibinfo{author}{\bibfnamefont{A.}~\bibnamefont{Furusawa}},
  \bibinfo{author}{\bibfnamefont{J.~L.} \bibnamefont{S\"{o}rensen}},
  \bibinfo{author}{\bibfnamefont{S.~L.} \bibnamefont{Braunstein}},
  \bibinfo{author}{\bibfnamefont{C.~A.} \bibnamefont{Fuchs}},
  \bibinfo{author}{\bibfnamefont{H.~J.} \bibnamefont{Kimble}},
  \bibnamefont{and} \bibinfo{author}{\bibnamefont{E.S.Polzik}},
  \bibinfo{journal}{Science} \textbf{\bibinfo{volume}{282}},
  \bibinfo{pages}{706} (\bibinfo{year}{1998}).

\bibitem[{\citenamefont{Grosshans et~al.}(2003)\citenamefont{Grosshans, Assche,
  Wenger, Brouri, Cerf, and Grangier}}]{grosshans:gaussianQKD}
\bibinfo{author}{\bibfnamefont{F.}~\bibnamefont{Grosshans}},
  \bibinfo{author}{\bibfnamefont{G.~V.} \bibnamefont{Assche}},
  \bibinfo{author}{\bibfnamefont{J.}~\bibnamefont{Wenger}},
  \bibinfo{author}{\bibfnamefont{R.}~\bibnamefont{Brouri}},
  \bibinfo{author}{\bibfnamefont{N.~J.} \bibnamefont{Cerf}}, \bibnamefont{and}
  \bibinfo{author}{\bibfnamefont{P.}~\bibnamefont{Grangier}},
  \bibinfo{journal}{Nature} \textbf{\bibinfo{volume}{421}},
  \bibinfo{pages}{238} (\bibinfo{year}{2003}),
  \urlprefix\url{http://www.nature.com/nature/journal/v421/n6920/abs/nature012%
89.html}.

\bibitem[{\citenamefont{Lance et~al.}(2004)\citenamefont{Lance, Symul, Bowen,
  Sanders, and Lam}}]{lance:2004}
\bibinfo{author}{\bibfnamefont{A.~M.} \bibnamefont{Lance}},
  \bibinfo{author}{\bibfnamefont{T.}~\bibnamefont{Symul}},
  \bibinfo{author}{\bibfnamefont{W.~P.} \bibnamefont{Bowen}},
  \bibinfo{author}{\bibfnamefont{B.~C.} \bibnamefont{Sanders}},
  \bibnamefont{and} \bibinfo{author}{\bibfnamefont{P.~K.} \bibnamefont{Lam}},
  \bibinfo{journal}{Physical Review Letters} \textbf{\bibinfo{volume}{92}},
  \bibinfo{pages}{177903} (\bibinfo{year}{2004}).

\bibitem[{\citenamefont{Takei et~al.}(2005)\citenamefont{Takei, Yonezawa, Aoki,
  and Furusawa}}]{takei:2005}
\bibinfo{author}{\bibfnamefont{N.}~\bibnamefont{Takei}},
  \bibinfo{author}{\bibfnamefont{H.}~\bibnamefont{Yonezawa}},
  \bibinfo{author}{\bibfnamefont{T.}~\bibnamefont{Aoki}}, \bibnamefont{and}
  \bibinfo{author}{\bibfnamefont{A.}~\bibnamefont{Furusawa}},
  \bibinfo{journal}{Physical Review Letters} \textbf{\bibinfo{volume}{94}},
  \bibinfo{pages}{220502} (\bibinfo{year}{2005}).

\bibitem[{\citenamefont{Munro et~al.}(2001)\citenamefont{Munro, James, White,
  and Kwiat}}]{munro:2001}
\bibinfo{author}{\bibfnamefont{W.~J.} \bibnamefont{Munro}},
  \bibinfo{author}{\bibfnamefont{D.~F.~V.} \bibnamefont{James}},
  \bibinfo{author}{\bibfnamefont{A.~G.} \bibnamefont{White}}, \bibnamefont{and}
  \bibinfo{author}{\bibfnamefont{P.~G.} \bibnamefont{Kwiat}},
  \bibinfo{journal}{Physical Review A} \textbf{\bibinfo{volume}{64}},
  \bibinfo{pages}{030302} (\bibinfo{year}{2001}).

\bibitem[{\citenamefont{Wei et~al.}(2005)\citenamefont{Wei, Altepeter,
  Branning, Goldbart, James, Jeffrey, Kwiat, Mukhopadhyay, and
  Peters}}]{wei:2005}
\bibinfo{author}{\bibfnamefont{T.-C.} \bibnamefont{Wei}},
  \bibinfo{author}{\bibfnamefont{J.~B.} \bibnamefont{Altepeter}},
  \bibinfo{author}{\bibfnamefont{D.}~\bibnamefont{Branning}},
  \bibinfo{author}{\bibfnamefont{P.~M.} \bibnamefont{Goldbart}},
  \bibinfo{author}{\bibfnamefont{D.~F.~V.} \bibnamefont{James}},
  \bibinfo{author}{\bibfnamefont{E.}~\bibnamefont{Jeffrey}},
  \bibinfo{author}{\bibfnamefont{P.~G.} \bibnamefont{Kwiat}},
  \bibinfo{author}{\bibfnamefont{S.}~\bibnamefont{Mukhopadhyay}},
  \bibnamefont{and} \bibinfo{author}{\bibfnamefont{N.~A.}
  \bibnamefont{Peters}}, \bibinfo{journal}{Physical Review A}
  \textbf{\bibinfo{volume}{71}}, \bibinfo{pages}{032329}
  (\bibinfo{year}{2005}).

\bibitem[{\citenamefont{Vidal and
  Werner}(2002)}]{vidal:computable_entanglement}
\bibinfo{author}{\bibfnamefont{G.}~\bibnamefont{Vidal}} \bibnamefont{and}
  \bibinfo{author}{\bibfnamefont{R.~F.} \bibnamefont{Werner}},
  \bibinfo{journal}{Physical Review A} \textbf{\bibinfo{volume}{65}},
  \bibinfo{pages}{032314} (\bibinfo{year}{2002}),
  \urlprefix\url{http://link.aps.org/abstract/PRA/v65/e032314}.

\bibitem[{\citenamefont{Peres}(1996)}]{peres:separability}
\bibinfo{author}{\bibfnamefont{A.}~\bibnamefont{Peres}},
  \bibinfo{journal}{Physical Review Letters} \textbf{\bibinfo{volume}{77}},
  \bibinfo{pages}{1413} (\bibinfo{year}{1996}),
  \urlprefix\url{http://link.aps.org/abstract/PRL/v77/p1413}.

\bibitem[{\citenamefont{Einstein et~al.}(1935)\citenamefont{Einstein, Podolsky,
  and Rosen}}]{einstein:EPR}
\bibinfo{author}{\bibfnamefont{A.}~\bibnamefont{Einstein}},
  \bibinfo{author}{\bibfnamefont{B.}~\bibnamefont{Podolsky}}, \bibnamefont{and}
  \bibinfo{author}{\bibfnamefont{N.}~\bibnamefont{Rosen}},
  \bibinfo{journal}{Physical Review} \textbf{\bibinfo{volume}{47}},
  \bibinfo{pages}{777} (\bibinfo{year}{1935}),
  \urlprefix\url{http://link.aps.org/abstract/PR/v47/p777}.

\bibitem[{\citenamefont{Braunstein and
  Kimble}(1998)}]{braunstein:contvarteleportation}
\bibinfo{author}{\bibfnamefont{S.}~\bibnamefont{Braunstein}} \bibnamefont{and}
  \bibinfo{author}{\bibfnamefont{H.~J.} \bibnamefont{Kimble}},
  \bibinfo{journal}{Physical Review Letters} \textbf{\bibinfo{volume}{80}},
  \bibinfo{pages}{869} (\bibinfo{year}{1998}),
  \urlprefix\url{http://link.aps.org/abstract/PRL/v80/p869}.

\bibitem[{\citenamefont{Mista et~al.}(2002)\citenamefont{Mista, Jr., Filip, and
  Fiurasek}}]{mista:cv_werner}
\bibinfo{author}{\bibfnamefont{L.}~\bibnamefont{Mista}},
  \bibinfo{author}{\bibnamefont{Jr.}},
  \bibinfo{author}{\bibfnamefont{R.}~\bibnamefont{Filip}}, \bibnamefont{and}
  \bibinfo{author}{\bibfnamefont{J.}~\bibnamefont{Fiurasek}},
  \bibinfo{journal}{Physical Review A} \textbf{\bibinfo{volume}{65}},
  \bibinfo{eid}{062315} (pages~\bibinfo{numpages}{8}) (\bibinfo{year}{2002}),
  \urlprefix\url{http://link.aps.org/abstract/PRA/v65/e062315}.

\bibitem[{\citenamefont{Cochrane et~al.}(2002)\citenamefont{Cochrane, Ralph,
  and Milburn}}]{cochrane:2002}
\bibinfo{author}{\bibfnamefont{P.~T.} \bibnamefont{Cochrane}},
  \bibinfo{author}{\bibfnamefont{T.~C.} \bibnamefont{Ralph}}, \bibnamefont{and}
  \bibinfo{author}{\bibfnamefont{G.~J.} \bibnamefont{Milburn}},
  \bibinfo{journal}{Physical Review A} \textbf{\bibinfo{volume}{65}},
  \bibinfo{pages}{062306} (\bibinfo{year}{2002}).

\bibitem[{\citenamefont{Grosshans and
  Cerf}(2004)}]{grosshans:gaussianQKDoptimal}
\bibinfo{author}{\bibfnamefont{F.}~\bibnamefont{Grosshans}} \bibnamefont{and}
  \bibinfo{author}{\bibfnamefont{N.~J.} \bibnamefont{Cerf}},
  \bibinfo{journal}{Physical Review Letters} \textbf{\bibinfo{volume}{92}},
  \bibinfo{pages}{047905} (\bibinfo{year}{2004}).

\end{thebibliography}

\end{document}